\documentclass[12pt]{iopart}

\usepackage{iopams,graphicx,amssymb}
\eqnobysec

\begin{document}

\title[Effects of turbulent mixing on the critical behaviour]
{Effects of turbulent mixing on critical
behaviour: Renormalization group analysis of the Potts model}

\author{ N V Antonov and A V Malyshev}

\address{Department of Theoretical Physics, St.~Petersburg State University
\\ Uljanovskaja 1, St.~Petersburg, Petrodvorez, 198504 Russia}

\ead{nikolai.antonov@pobox.spbu.ru, alvlamal@gmail.com}

\begin{abstract}
Critical behaviour of a system, subjected to strongly anisotropic turbulent
mixing, is studied by means of the field theoretic renormalization group.
Specifically, relaxational stochastic dynamics of a non-conserved
multicomponent order parameter of the Ashkin--Teller--Potts model, coupled
to a random velocity field with prescribed statistics, is considered.
The velocity is taken Gaussian, white in time, with correlation
function of the form $\propto \delta(t-t') /|{\bf k}_{\bot}|^{d-1+\xi}$,
where ${\bf k}_{\bot}$ is the component of the wave vector, perpendicular
to the distinguished direction (``direction of the flow'') --- the
$d$-dimensional generalization of the ensemble introduced by
Avellaneda and Majda [1990 {\it Commun. Math. Phys.} {\bf 131} 381]
within the context of passive scalar advection. This model can describe
a rich class of physical situations. It is shown that, depending on
the values of parameters that define self-interaction of the order parameter
and the relation between the exponent $\xi$ and the space dimension $d$,
the system exhibits various types of large-scale scaling behaviour,
associated with different infrared attractive fixed points of the
renormalization-group equations. In addition to known asymptotic regimes
(critical dynamics of the Potts model and passively advected field without
self-interaction), existence of a new, non-equilibrium and strongly
anisotropic, type of critical behaviour (universality class) is established,
and the corresponding critical dimensions are calculated to the leading
order of the double expansion in $\xi$ and $\varepsilon=6-d$ (one-loop
approximation). The scaling appears strongly anisotropic in the sense that
the critical dimensions related to the directions parallel and perpendicular
to the flow are essentially different.
\end{abstract}

\pacs{05.10.Cc, 05.70.Jk, 64.60.ae, 64.60.Ht, 47.27.ef}

\maketitle

\section{Introduction} \label{sec:Intro}

Numerous systems of very different physical nature reveal interesting
singular behaviour in the vicinity of their critical points. Their
correlation functions exhibit self-similar (scaling) behaviour with
universal critical dimensions: they depend only on few global characteristics
of the system (like symmetry or space dimension). Consistent qualitative
and quantitative description of the
critical behaviour is provided by the field theoretic
renormalization group (RG). In the RG approach, possible types of critical
regimes (universality classes) are associated with infrared (IR) attractive
fixed points of renormalizable field theoretic models.

Most typical equilibrium phase transitions belong to the universality class
of the $O_{n}$-symmetric $\varphi^{4}$ model of an $n$-component scalar
order parameter $\varphi$. Universal characteristics of the
critical behaviour depend only on $n$ and the space dimension $d$ and can
be calculated within systematic perturbation schemes, in particular,
in the form of expansion in $\varepsilon=4-d$, the deviation of the
space dimension from its upper critical value $d=4$;
see the monographs \cite{Zinn,Book3} and the literature cited therein.

Another important example is provided by the Ashkin--Teller--Potts (ATP)
class of models \cite{ATP}--\cite{Bonfim}. In the continuous formulation,
they are described by the effective Hamiltonian for the $n$-component
order parameter with a trilinear interaction term, invariant under the
hypertetrahedron symmetry group \cite{Golner}--\cite{Bonfim}. Such models
have numerous physical applications: magnetic materials and solids with
nontrivial symmetry, Edwards-Anderson spin-glass models within the
replica formalism \cite{EdA}, and so on. In general, the ATP models
describe systems which locally have $n$ states, but the energy of any
given configuration depends on whether pairs of neighboring sites
are in the same state or not \cite{ATP}.
The case $n=2$ corresponds to nematic-to-isotropic transitions
in the liquid crystals \cite{DeGe}, while the formal limits
$n=0$ and $n=-1$ correspond to the percolation problem and the
random resistor network, respectively \cite{FaS}--\cite{Ha}.
Recently, models with trilinear interaction have attracted a new amount
of interest due to their interesting formal properties
\cite{Janssen,Bender}
and applications to the dynamics of first-order phase transitions
\cite{Chen}. The application of the cubic model to the Yang-Lee edge
singularity has long been known \cite{YL}.

The problem of the nature of the phase transition in the ATP model has a
long and rather entangled history; see {\it e.g.}
\cite{Baxter}--\cite{Amit} and references
therein. According to Landau's phenomenological theory, existence of a
trilinear term excludes the possibility of the second-order transition.
On the contrary, exact two-dimensional results, numerical simulations and
RG analysis suggest that for small $n$, the phase transition in the ATP
model is of the second order, while for $n$ large enough ($n>3$ in two
dimensions \cite{Baxter} and $n>7/3$ in the vicinity of the upper critical
dimension $d=6$ \cite{Priest,Amit}) the transition becomes a first-order one.
In this paper, we accept the point of view that the existence of an IR
attractive fixed point of the RG equations implies the existence of a
self-similar (scaling) asymptotic regime and thus the existence of a kind
of critical state.

It is well known that
dynamical critical behaviour (critical singularities of relaxation and
correlation times, various kinetic and transport coefficients {\it etc.})
appears much richer, less universal and is comparatively less understood.
Different nature of the order parameter (conserved or non-conserved),
inclusion of additional slow modes (densities of entropy or energy) and
interaction with hydrodynamical degrees of freedom produce different types
of critical dynamics for the same static model \cite{Book3,HH}.

The behaviour of a real system near its critical point is extremely
sensitive to external disturbances, gravity, geometry of the experimental
setup, presence of impurities and so on; see the monograph \cite{Ivanov}
for the general discussion and the references.
``Ideal'' equilibrium critical behaviour of an infinite system can be
obscured by limited accuracy of measuring the temperature, finite-size
effects, finite time of evolution (ageing) and so on.
In the presence of a distinguished direction, scaling behaviour of such
systems can become strongly anisotropic, with
different critical dimensions corresponding to different spatial directions.
What is more, some
disturbances (randomly distributed impurities in magnets and turbulent
mixing of fluid systems) can change the type of the phase transition
(first-order to second order one, and vice versa) and produce
completely new types of critical behaviour (universality classes)
with rich and rather exotic properties.

Investigation of the effects of various kinds of deterministic or chaotic
flows (laminar shear flows, turbulent convection and so on) on the behaviour
of the critical fluids (like liquid crystals or binary mixtures near their
consolution points) has shown that the flow can destroy the usual critical
behaviour: it can change to the mean-field behaviour or, under some
conditions, to a more complex behaviour described by new non-equilibrium
universality classes \cite{Beysens}--\cite{AntM}.

In this paper we study effects of turbulent mixing on the dynamical critical
behaviour of the systems, described by the generalized ATP model, paying
special attention to anisotropy of the flow. Bearing in mind application to
liquid crystals or percolation in liquid media, we consider a purely
relaxational stochastic dynamics of a non-conserved order parameter of
the ATP model, coupled to a random velocity field with prescribed Gaussian
statistics.

Recently, the models involving passive scalar fields advected by such
synthetic velocity ensembles attracted a great deal of attention
because of the insight they offer into the origin of intermittency and
anomalous scaling in the real fluid turbulence; see the review paper
\cite{FGV} and references therein. The RG approach to the problem of
passive advection is reviewed in \cite{JphysA}.
In spite of their relative simplicity,
such models reproduce many of the anomalous features of genuine turbulent
heat or mass transport observed in experiments. In the context of our study,
it is especially important that they allow to easily model anisotropy of
the flow, which in more realistic models would be introduced by the initial
and/or boundary conditions. More specifically, we employ the
$d$-dimensional generalization of a strongly anisotropic ensemble
introduced in \cite{AM} and further discussed in a number of papers,
{\it e.g.} \cite{Glimm}--\cite{shark}, in connection with the passive scalar
problem: the velocity field is oriented along a chosen direction
$\bf n$ and its correlation function depends only on the coordinates
perpendicular to $\bf n$.

Although simplified, the model appears rather nontrivial and captures
the main property of the problem: existence of a new, non-equilibrium and
strongly anisotropic, universality class of scaling behaviour.

The plan of the paper is as follows. In section~\ref{sec:Models} we present
the detailed description of the model and its field theoretic formulation.
In section~\ref{sec:Reno} we analyze the ultraviolet (UV) divergences,
relaying upon the power counting and additional symmetry considerations.
We show that the model, after proper extension,
appears multiplicatively renormalizable. Thus we can derive the RG equations
and introduce the RG functions ($\beta$ functions and anomalous dimensions
$\gamma$) in the standard manner; see section~\ref{sec:RGE}.

In section~\ref{sec:FPS} we show that, depending on the relation between
the spatial dimension $d$ and the exponent $\xi$ in the velocity correlator,
the model reveals four different types of critical behaviour, associated
with four fixed points of the corresponding RG equations. Three fixed
points correspond to known regimes: Gaussian or free field theory,
non-interacting scalar field passively advected by the flow (the ATP
nonlinearity in the original dynamical equations appears irrelevant),
and the original critical behaviour of the model without mixing.
The most interesting fourth point corresponds to a new full-scale
non-equilibrium universality class, in which both the nonlinearity and
turbulent mixing are relevant.

The corresponding critical dimensions can
be calculated as double expansions in two parameters: $\xi$ and
$\varepsilon=4-d$. The scaling behaviour appears strongly anisotropic
in the sense that the critical dimensions related to the directions
parallel and perpendicular to the flow are essentially different.
The practical calculation of the renormalization constants, RG functions,
regions of stability and critical dimensions was performed in the
leading order (one-loop approximation); some of the results, however, are
exact (valid to all orders of the double $\varepsilon$--$\xi$ expansion).
These issues are discussed in section ~\ref{sec:DimeNS}, while
section~\ref{sec:Conc} is reserved for conclusion.

\section{Description of the model and the field theoretic formulation}
\label{sec:Models}

Relaxational dynamics of a non-conserved $n$-component order parameter
$\varphi_a(x)$ with $x\equiv\{t,\textbf{x}\}$ is described by a
stochastic differential equation
\begin{eqnarray}
\partial_{t} \varphi_a(x) =-\lambda_0\frac{\delta {\cal H}(\varphi)}{
\delta \varphi_a(x)} + \eta_a(x),
\label{eq1}
\end{eqnarray}
where $\partial_{t} = \partial/ \partial {t}$,  $\lambda_0$ is the (constant) kinetic
coefficient,  and $\eta_a(x)$ is a Gaussian random noise
with zero mean and the pair correlation function
\begin{eqnarray}
\langle \eta_a(x)\eta_b(x') \rangle = \delta_{ab} D_{\eta} \,(x-x'),
\nonumber \\
D_{\eta} \,(x-x') \equiv 2\lambda_0\,\delta(t-t')\,
\delta^{(d)}(\bf x-\bf x') ,
\label{forceD}
\end{eqnarray}
$d$ being the dimension of the $\bf x$ space.
Near the critical point, the Hamiltonian ${\cal H}(\varphi)$ of
the ATP model is taken in the form \cite{Golner,Zia,Priest}
\begin{eqnarray}
{\cal H}(\varphi) &=&  \int d{\bf x} \left\{- \frac{1}{2}\,
 \varphi_a({\bf x})\partial^{2} \varphi_a({\bf x}) + \frac{\tau_{0}}{2}\,
\varphi_a({\bf x})\varphi_a({\bf x}) + \right.
\nonumber \\
&+& \left. \frac{g_{0}}{3!}R_{abc} \,
\varphi_a({\bf x})\varphi_b({\bf x})\varphi_c({\bf x}) \right\},
\label{LG}
\end{eqnarray}
where $\partial_{i} = \partial/ \partial x_{i}$ is the spatial derivative,
$\partial^{2} = \partial_{i}\partial_{i}$ is the Laplacian,
$\tau_{0} \propto (T-T_{c})$ measures deviation of the temperature
(or its analog) from the critical value and $g_0$ is the coupling constant.
Summations over repeated indices are always implied
($a,b,c=1,\dots,n$ and $i=1,\dots,d$);
after the functional differentiation in (\ref{eq1}) one has to replace
$\varphi({\bf x}) \to \varphi(x)$.

Following \cite{Bonfim}, we consider the generalized case of certain symmetry
group ${\cal G}$, which has the only irreducible invariant third-rank tensor
$R_{abc}$; with no loss of generality it is assumed to be symmetric.
In the original ATP model ${\cal G}$ is the symmetry group of the
hypertetrahedron in $n$ dimensions. Then the tensor $R_{abc}$
is conveniently expressed in terms of the set of $n+1$ vectors $e^{\alpha}$
which define its vertices \cite{Golner,Zia}:
\begin{eqnarray}
R_{abc}=\sum_{\alpha}e^{\alpha}_ae^{\alpha}_be^{\alpha}_c, \nonumber
\label{tensor}
\end{eqnarray}
where the $e^{\alpha}_a$ satisfy
\begin{eqnarray}
\sum_{\alpha=1}^{n+1}e^{\alpha}_a=0, \;\;
\sum_{\alpha=1}^{n+1}e^{\alpha}_ae^{\alpha}_b=(n+1)\delta_{ab}, \;\;
\sum_{a=1}^{n} e^{\alpha}_ae^{\beta}_a=(n+1)\delta^{\alpha\beta}-1.
\label{tensor2}
\end{eqnarray}
Using equations (\ref{tensor2}) all the contractions with the tensor
$R_{abc}$ can be calculated. For example, the following contractions
of two and three tensors have the form
\begin{eqnarray}
R_{abc}R_{abe}=R_1 \delta_{ce}, \quad
R_{aec}R_{chb}R_{b\!f\!a}=R_2 R_{eh\!f},
\label{rr}
\end{eqnarray}
where
\begin{eqnarray}
 R_1=(n+1)^2(n-1), \quad  R_2=(n+1)^2(n-2).
\label{rr2}
\end{eqnarray}

Coupling with the velocity field ${\bf v}= \{v_{i}(x)\}$ is introduced
by the replacement
\begin{eqnarray}
\partial_{t} \to \nabla_{t} = \partial_{t} + v_{i} \partial_{i},
\label{nabla}
\end{eqnarray}
where $\nabla_{t}$ is the Lagrangian (Galilean covariant) derivative.
For incompressible fluid, the velocity
field ${\bf v}$ is transverse due to the continuity
relation: $\partial_{i} v_{i}=0$. The velocity ensemble is defined as
follows \cite{AM}.
Let ${\bf n}$ be a unit constant vector that determines some distinguished
direction (`direction of the flow'). Then any vector can be decomposed
into the components perpendicular and parallel to the flow, for example,
${\bf x} = {\bf x}_{\bot} + {\bf n}\, x_{\parallel}$ with
${\bf x}_{\bot} \cdot {\bf n} =0$.
The velocity field will be taken in the form
\begin{eqnarray}
{\bf v} = {\bf n} v(t, {\bf x}_{\bot}),
\label{vello} \nonumber
\end{eqnarray}
where $v(t, {\bf x}_{\bot})$ is a scalar function independent of
$x_{\parallel}$.
Then the incompressibility condition is automatically satisfied:
\begin{eqnarray}
\partial_{i} v_{i} = \partial_{\parallel} v(t, {\bf x}_{\bot}) = 0.
\label{inko} \nonumber
\end{eqnarray}

For $v(t, {\bf x}_{\bot})$ we assume a Gaussian distribution with zero
mean and the pair correlation function of the form:
\begin{eqnarray}
\langle v(t,{\bf x}_{\bot}) v(t', {\bf x}_{\bot}') \rangle =
\delta(t-t') \int \frac{d {\bf k}}{(2\pi)^{d}} \,
\exp \left\{ {\rm i} {\bf k}\cdot ({\bf x}-{\bf x}') \right\} D_{v} (k)=
\nonumber \\
= \delta(t-t') \int \frac{d {\bf k}_{\bot}}{(2\pi)^{d-1}} \, \exp
\left\{ {\rm i} {\bf k}_{\bot}\cdot ({\bf x}_{\bot}-{\bf x}'_{\bot})
\right\} \widetilde D_{v} (k_{\bot}) , \quad  k_{\bot}=|{\bf k}_{\bot}|
\label{veloc1}
\end{eqnarray}
with the scalar coefficient functions 
\begin{eqnarray}
D_{v} (k)= 2\pi \delta(k_{\parallel}) \, \widetilde D_{v} (k_{\bot}) ,
\quad \widetilde D_{v} (k_{\bot}) = D_{0}\, k_{\bot}^{-d+1-\xi}.
\label{veloc2}
\end{eqnarray}
Here and below $d$ is the dimension of the ${\bf x}$ space, $D_{0}>0$
is a constant amplitude factor and $\xi$ an arbitrary exponent.
The IR regularization in (\ref{veloc1})
is provided by the cutoff $k_{\bot}>m$ (by dimension, $\tau_{0} \propto
m^{2}$). Precise form of the IR regularization is inessential;
sharp cutoff is the most convenient choice from the calculational
viewpoints. The natural interval for the
exponent is $0< \xi <2 $, when the so-called effective eddy diffusivity
has a finite limit for $m\to0$; it includes the most realistic Kolmogorov
value $\xi=4/3$.

In order to ensure multiplicative renormalizability of the model, it is
necessary to split the Laplacian in (\ref{eq1}) into the parallel and
perpendicular parts $\partial^{2} \to \partial^{2}_{\bot}
+ f_{0} \partial^{2}_{\parallel}$ by introducing a new parameter
$f_{0}>0$. Here $\partial^{2}_{\bot}$ is the Laplacian in the subspace
orthogonal to the vector ${\bf n}$ and
$ \partial_{\parallel} = \partial/ \partial x_{\parallel}$.
In the anisotropic case, these two terms will be renormalized
in a different way; more detailed discussion of this point can be found
in \cite{Alexa,AntM}. Thus equation (\ref{eq1}) becomes
\begin{eqnarray}
\nabla_{t} \varphi_a =\lambda_0 \left(\partial^{2}_{\bot} + f_{0}
\partial^{2}_{\parallel}-\tau_0\right)\varphi_a -
\frac{g_0\lambda_0}{2}\,R_{abc}\,\varphi_b\varphi_c+\eta_a;
\label{eq2}
\end{eqnarray}
this completes formulation of the model.

According to the general rule (see {\it e.g.} chap.~5 of the monograph
\cite{Book3}), our stochastic problem is equivalent to the field theoretic
model of the extended set of fields
$\Phi = \left\{ \varphi^{\prime}_a, \varphi_a, {\bf v} \right\}$
with the action functional
\begin{eqnarray}
{\cal S} (\Phi) = \frac{ \varphi^{\prime}_a D_{\eta} \varphi^{\prime}_a }{2}
+ \varphi^{\prime}_a  \left\{ - \nabla_{t}  +  \lambda_0
\left(  \partial^{2}_{\bot} + f_{0}
\partial^{2}_{\parallel} \right) -\lambda_0 \tau_0 \right\}\varphi_a-
\nonumber\\
-\frac{\lambda_0 g_0 f^{1/4}_0}{2}\,R_{abc}\,\varphi^{\prime}_a
\varphi_b\varphi_c + {\cal S}_{v} ({\bf v}),
\label{action}
\end{eqnarray}
where we segregated the factor $f^{1/4}_0$ from the charge $g_0$.
The first few terms
represent the De Dominicis--Janssen action functional for the stochastic
problem (\ref{eq1}), (\ref{forceD}) at fixed ${\bf v}$; it involves the
auxiliary scalar response field $\varphi^{\prime}_a (x)$.
All the required
integrations over $x=\{t,{\bf x}\}$ and
summations over the vector indices are implied, for example,
\[ \varphi^{\prime}_a  \partial^{2}_{\bot} \varphi_a =
\sum_{a=1}^n \int dt \int d{\bf x} \,
\varphi^{\prime}_a (x) \partial^{2}_{\bot} \varphi_a(x).  \]

The last term in (\ref{action}) corresponds to the Gaussian averaging over
${\bf v}$ with the correlator (\ref{veloc1}) and has the form
\begin{eqnarray}
{\cal S}_{v} ( {\bf v}) = - \frac{1}{2}\,
\int dt \int d{\bf x}_{\bot} d{\bf x}_{\bot}' v(t,{\bf x}_{\bot})
\widetilde D^{-1}_{v} ({\bf x}_{\bot}-{\bf x}'_{\bot}) v(t,{\bf x}_{\bot}'),
\label{Sv} \nonumber
\end{eqnarray}
where
\begin{eqnarray}
\widetilde D^{-1}_{v} ({\bf r}_{\bot}) \propto D_{0}^{-1}
\, r_{\bot}^{2(1-d)-\xi}
\label{Dv} \nonumber
\end{eqnarray}
is the kernel of the inverse linear operation $D^{-1}_{v}$ for the
correlation function $D_{v}$ in (\ref{veloc2}).

This formulation means that statistical averages of random quantities
in the original stochastic problem coincide with the Green functions of the
field theoretic model with the action (\ref{action}), given by functional
averages with the weight $\exp {\cal S}(\Phi)$. This allows one to apply
the standard Feynman diagrammatic technique, the field theoretic
renormalization theory and renormalization group to our stochastic problem.

\section{Canonical dimensions, UV divergences and renormalization}
\label{sec:Reno}

It is well known that the analysis of UV divergences is based on the analysis
of canonical dimensions; see {\it e.g.} \cite{Zinn,Book3}. In general,
dynamic models have two scales: canonical dimension of some
quantity $F$ (a field or a parameter in the action functional) is completely
characterized by two numbers, the frequency dimension $d_{F}^{\omega}$
and the momentum dimension $d_{F}^{k}$; see {\it e.g.} chap.~5 in
\cite{Book3}. They are determined such that
$[F] \sim [T]^{-d_{F}^{\omega}} [L]^{-d_{F}^{k}}$, where $L$ is some
length scale and $T$ is the time scale.

Our strongly anisotropic model, however, has two independent momentum
scales, related to the directions perpendicular and parallel to the
vector ${\bf n}$, and requires a more detailed specification of the
canonical dimensions. Namely, one has to introduce two independent momentum
canonical dimensions $d_{F}^{\bot}$ and $d_{F}^{\parallel}$ so that
\[ [F] \sim [T]^{-d_{F}^{\omega}}  [L_{\bot}]^{-d_{F}^{\bot}}
[L_{\parallel}]^{-d_{F}^{\parallel}}, \]
where $L_{\bot}$ and $L_{\parallel}$ are (independent) length scales in the
corresponding subspaces. The dimensions are found from the obvious
normalization conditions $d_{k_{\bot}}^{\bot}= -d_{\bf x_{\bot}}^{\bot}=1$,
$d_{k_{\bot}}^{\parallel}=-d_{\bf x_{\bot}}^{\parallel}=0$,
$d_{k_{\bot}}^{\omega} = d_{k_{\parallel}}^{\omega}=0$,
$d_{\omega }^{\omega }=-d_t^{\omega }=1$, and so on, and from the
requirement that each term of the action functional (\ref{action})
be dimensionless (with respect to all the three independent dimensions
separately).

\begin{table}
\caption{Canonical dimensions of the fields and parameters
in the model (\protect\ref{action})}
\label{table1}
\begin{tabular}{|c|c|c|c|c|c|c|c|c|c|}
\hline
$F$ & $\varphi^{\prime}_a  $ & $\varphi_a$ & $ {\bf v} $ &
$\lambda_0, \lambda$ &
$f_0 ,f$ & $\tau_0, \tau$& $\mu$ & $g_0$  & $w_0$ \\
\hline
$d_{F}^{\omega}$ & 0 & 0 & 1 & 1 & 0 & 0&  0 &  0 &  0 \\
\hline
$d_{F}^{\parallel}$ & 1/2 & 1/2 & $-1$ & 0 & $-2$ & 0 &  0 & 0  & 0 \\
\hline
$d_{F}^{\bot}$ & $(d+1)/2$ & $(d-3)/2$ & 0 & $-2$ & 2 & 2 &  1&
$(6-d)/2$&  $\xi$  \\
\hline
\end{tabular}
\end{table}
The canonical dimensions of the model (\ref{action}) are given in
table~\ref{table1}, including renormalized parameters, which will be
introduced a bit later. From table~\ref{table1} it follows that the
model is logarithmic (the coupling constants $g_{0}$ and $w_{0}$ are
simultaneously dimensionless) at $d=6$ and $\xi=0$, so that the UV
divergences in the correlation functions manifest themselves as poles
in $\varepsilon \equiv 6-d$, $\xi$ and their linear combinations.

The total canonical dimension can be found from the relation
$d_{F} = d_{F}^{\bot} + d_{F}^{\parallel} +2d_{F}^{\omega}$
(in the free theory,
$\partial_{t}\propto\partial^{2}_{\bot} \propto \partial^{2}_{\parallel}$).
In the renormalization theory it plays the same part as the conventional
(momentum) dimension does in one-scale static problems: superficial UV
divergences, whose removal requires counterterms, can be present only in
those 1-irreducible Green functions, for which the total canonical dimension
at the logarithmic values $\varepsilon =\xi=0$ (formal index of divergence)
is a non-negative integer.

The careful analysis of the table~\ref{table1}, augmented by symmetry
considerations,
shows that all the counterterms needed to cancel the UV divergences in
our model are present in the action (\ref{action}).
Here, important role is played by the Galilean symmetry and
the invariance  with respect to the group ${\cal G}$.
For example, the function
$\left\langle \varphi^{\prime}\varphi  vv\right\rangle$ can be
omitted from consideration
because the corresponding counterterm $\varphi^{\prime}_a\varphi_a  v^2$
is forbidden by the Galilean invariance. A similar situation
occurs for the functions $\left\langle \varphi^{\prime}  v\right\rangle$
and $\left\langle \varphi^{\prime} vv\right\rangle$.
For the first function, a counterterm necessarily
contains a spatial derivative, whose vector index is contracted
with the velocity  field; this gives zero in view of the
transversality condition (\ref{inko}). For the second function, the
possible counterterms
$\varphi^{\prime}_a  (\partial_iv_k)(\partial_kv_i)$ and
$\varphi^{\prime}_a  (\partial_iv_k)(\partial_i v_k)$
are forbidden by the symmetry with respect to ${\cal G}$.
Furthermore, the Galilean symmetry requires that the counterterms
$\varphi^{\prime}_a  \partial _{t}\varphi_a$ and
$\varphi^{\prime}_a  (v_{i}\partial_{i}) \varphi_a$
enter the renormalized action only in the form of the Lagrangian derivative
$\varphi^{\prime}_a  \nabla _{t}\varphi_a$.
In turn, owing to the symmetry with respect to ${\cal G}$, the
trilinear term in (\ref{action}) is renormalized as a single entity.

Thus our model appears multiplicatively renormalizable with
the renormalized action of the form
\begin{eqnarray}
{\cal S}_{R} (\Phi) = Z_{1}\frac{ \varphi^{\prime}_a
D_{\eta} \varphi^{\prime}_a }{2} +
\varphi^{\prime}_a  \left\{ -Z_{2} \nabla_{t}  +  \lambda
\left( Z_{3} \partial^{2}_{\bot} + Z_{4}f
\partial^{2}_{\parallel} \right) -Z_{5}\lambda \tau \right\}\varphi_a-
\nonumber\\
-Z_{6}\frac {\lambda g \mu^{\varepsilon/2}f^{1/4}}{2}\,R_{abc}\,
\varphi^{\prime}_a \varphi_b\varphi_c + {\cal S}_{v} ({\bf v}).
\label{RenAct}
\end{eqnarray}
Here $\lambda$, $\tau$, $f$, $g$ and $w$ are renormalized analogs of the
bare parameters (with the subscripts ``0'') and $\mu$ is the reference mass
scale (additional arbitrary parameter of the renormalized theory). The
renormalization constants capture all the divergences at $\varepsilon$,
$\xi=0$ so that the correlation functions of the renormalized model
(\ref{RenAct}) have finite limits for $\varepsilon$, $\xi\to 0$, when
expressed in renormalized parameters $\lambda$, $\mu$ and so on.

Simple analysis shows that the Feynman diagrams needed for the calculation of
the renormalization constants in (\ref{RenAct}) are the same as in the
one-component $\varphi^3$ theory (model 2 from \cite{AntM}), multiplied
with the appropriate tensor contractions.
Moreover, in the one-loop approximation only contractions (\ref{rr}) appear.
Therefore, one can easily generalize the results of \cite{AntM} to the case
at hand with the aid of (\ref{rr}).

In \cite{AntM} the minimal subtraction (MS) renormalization scheme was
employed. In the MS scheme the renormalization constants have the forms
$Z_{i}=1+\,$ only singularities in $\varepsilon$ and $\xi$, with the
coefficients depending on the two completely dimensionless parameters ---
renormalized coupling constants $g$ and $w$. The one-loop results
for the renormalization constants in (\ref{RenAct}) are the following
\begin{eqnarray}
Z_{1}=Z_{2}=1-\frac{uR_1}{2\varepsilon}, \quad
Z_{3}=1-\frac{uR_1}{3\varepsilon}, \quad
Z_{4}=1-\frac{uR_1}{3\varepsilon}-\frac{w}{\xi}, \nonumber \\
Z_{5}=1-\frac{2uR_1}{\varepsilon}, \quad
Z_{6}=1-\frac{2uR_2}{\varepsilon} .
\label{Zz}
\end{eqnarray}
Here we have passed to more convenient coupling constants
$u\rightarrow g^2/128\pi^3$ and $w\rightarrow w/24\pi^3$.

The parameters $R_1$ and $R_2$ are related to the dimension $n$ of the order
parameter by the expression (\ref{rr2}). Although we are especially
interested in the cases $n=0$ and $2$, for completeness the
coefficients $R_1$ and $R_2$ in what follows are assumed to be
arbitrary.

Expression (\ref{RenAct}) is equivalent to the multiplicative
renormalization of the fields $\varphi_a \to \varphi_a Z_{\varphi}$,
$\varphi^{\prime}_a \to \varphi^{\prime}_a Z_{\varphi'}$ and the parameters:
\begin{eqnarray}
\lambda_{0} &=& \lambda Z_{\lambda}, \;\; \tau_{0} = \tau Z_{\tau}, \;\;
f_{0} = f Z_{f}, \nonumber \\
g_{0} &=& g \mu^{\varepsilon/2} Z_{g}\,
\left(u_{0} = u \mu^{\varepsilon} Z_{u}\right),  \;\;
w_{0} = w \mu^{\xi} Z_{w}
\label{Multy}
\end{eqnarray}
(no renormalization of the velocity field is needed: $Z_{v}=1$).
The constants in Eqs. (\ref{RenAct}) and (\ref{Multy}) are related
as follows:
\begin{eqnarray}
Z_{1} = Z_{\lambda} Z^{2}_{\varphi'}, \quad
Z_{2} = Z_{\varphi'} Z_{\varphi}, \quad
Z_{3} = Z_{\varphi'} Z_{\lambda} Z_{\varphi}, \nonumber \\
Z_{4} = Z_{\varphi'} Z_{\lambda} Z_{f} Z_{\varphi}, \quad
Z_{5} = Z_{\varphi'} Z_{\lambda}  Z_{\tau} Z_{\varphi}, \quad
Z_{6} = Z_{\lambda} Z_{u}^{1/2} Z_{f}^{1/4} Z_{\varphi'} Z_{\varphi}^{2}.
\label{ZZ}
\end{eqnarray}
Since the last term ${\cal S}_{v} (\bf v)$ is not renormalized,
the amplitude $D_{0}$ is expressed in renormalized parameters as
\begin{eqnarray}
D_{0} = w_{0} f_{0} \lambda_{0} = w f\mu^{\xi}\lambda,
\label{RenD}
\nonumber
\end{eqnarray}
which leads to the relation
\begin{eqnarray}
Z_{w}Z_{f}Z_{\lambda}=1
\label{ZZ1}
\end{eqnarray}
between the renormalization constants.

\section{RG equations and RG functions} \label{sec:RGE}

Let us recall an elementary derivation of the RG equations; more detailed
discussion can be found in monographs \cite{Zinn,Book3}.
The RG equations are written for the renormalized
correlation functions $G_{R} =\langle \Phi\cdots\Phi\rangle_{R}$, which
differ from the original (unrenormalized) ones
$G =\langle \Phi\cdots\Phi\rangle$ only by normalization and choice of
parameters, and therefore can equally be used for analyzing the critical
behaviour. The relation
${\cal S}_{R} (Z_{\Phi} \Phi,e,\mu) = {\cal S}(\Phi,e_{0})$
between the
functionals (\ref{action}) and (\ref{RenAct}) results in the relations
\begin{equation}
G(e_{0},\dots) =  Z_{\varphi'}^{ N_{\varphi'}}
Z_{\varphi}^{\raisebox{1pt}{$\scriptstyle N_{\varphi}$}}
G_{R}(e,\mu,\dots)
\label{multi}
\end{equation}
between the correlation functions. Here,
$N_{\varphi'}$ and $N_{\varphi}$ are the full numbers of corresponding fields
entering into $\Gamma$ (we recall that in our model $Z_{v}=1$);
$e_{0}=\{\lambda_{0}, \tau_{0}, f_{0}, u_{0}, w_{0} \}$ is the full set of
bare parameters and $e=\{ \lambda, \tau, f, u, w \}$ are their renormalized
counterparts; the ellipsis stands for the other arguments
(times, coordinates, momenta {\it etc.}).

We use $\widetilde{\cal D}_{\mu}$ to denote the differential operation
$\mu\partial_{\mu}$ for fixed $e_{0}$ and operate on both sides of the
equation (\ref{multi}) with it. This gives the basic RG differential
equation:
\begin{equation}
\left\{ {\cal D}_{RG} + N_{\varphi}\gamma_{\varphi} +
N_{\varphi'}\gamma_{\varphi'} \right\} \,G^{R}(e,\mu,\dots) = 0,
\label{RG1}
\end{equation}
where ${\cal D}_{RG}$ is the operation $\widetilde{\cal D}_{\mu}$
expressed in the renormalized variables:
\begin{equation}
{\cal D}_{RG}\equiv {\cal D}_{\mu} + \beta_{u}\partial_{u} +
\beta_{w}\partial_{w} - \gamma_{f}{\cal D}_{f} -
\gamma_{\lambda}{\cal D}_{\lambda} - \gamma_{\tau}{\cal D}_{\tau}.
\label{RG2}
\end{equation}
Here we have written ${\cal D}_{x}\equiv x\partial_{x}$ for any variable
$x$, the anomalous dimensions $\gamma$ are defined as
\begin{equation}
\gamma_{F}\equiv \widetilde{\cal D}_{\mu} \ln Z_{F} \quad
{\rm for\ any\ quantity} \ F,
\label{RGF1}
\end{equation}
and the $\beta$ functions for the two dimensionless couplings $u$ and $w$ are
\begin{equation}
\beta_{u} \equiv \widetilde {\cal D}_{\mu} u = u\,[-\varepsilon-\gamma_{u}],
\quad
\beta_{w} \equiv \widetilde {\cal D}_{\mu} w = w\,[-\xi-\gamma_{w}],
\label{betagw}
\end{equation}
where the second equalities come from the definitions and the
relations (\ref{Multy}).

Equations (\ref{ZZ}) result in the following relations between the anomalous
dimensions
\begin{eqnarray}
\gamma_{1} = \gamma_{\lambda} + 2 \gamma_{\varphi'} , \quad
\gamma_{2} = \gamma_{\varphi'}+  \gamma_{\varphi} , \quad
\gamma_{3} = \gamma_{\varphi'} +\gamma_{\lambda}+  \gamma_{\varphi} , \nonumber \\
\gamma_{4} = \gamma_{f} +  \gamma_{3} , \quad \gamma_{5} =
\gamma_{\tau} +  \gamma_{3} , \quad \gamma_{6} =
\gamma_{\lambda}+\gamma_{u}/2+\gamma_{f}/4+\gamma_{\varphi'}+ 2\gamma_{\varphi} ,
\label{gammas}
\end{eqnarray}
while from (\ref{ZZ1}) one obtains
\begin{eqnarray}
\gamma_{f} + \gamma_{w} +\gamma_{\lambda}=0.
\label{gaas}
\end{eqnarray}
The dimensions $\gamma_{1}$--$\gamma_{6}$ are calculated from the
corresponding renormalization constants using the definition (\ref{RGF1}):
\begin{eqnarray}
\gamma_{1} =\gamma_{2} = \frac{uR_1}{2}, \quad
\gamma_{3} = \frac{uR_1}{3}, \quad
\gamma_{4} = \frac{uR_1}{3}+w,
\nonumber \\
\gamma_{5} = 2uR_1, \quad
\gamma_{6} = 2uR_2,
\label{gammaTwo}
\end{eqnarray}
with the corrections of the order $u^2$, $w^2$, $uw$ and higher.

The RG functions entering the equation (\ref{RG2}) are easily found from
the relations (\ref{Zz}), (\ref{gammas}) and (\ref{gaas}):
\begin{eqnarray}
\gamma_{\varphi'}= \gamma_{2}-\gamma_{3}/2=uR_1/3, \quad
\gamma_{\varphi}= \gamma_{3}/2=\frac{uR_1}{6}, \nonumber \\
\gamma_{f}=\gamma_{4}-\gamma_{3}=w, \quad
\gamma_{\tau} = \gamma_{5}-\gamma_{3}= 5uR_1/3, \nonumber \\
\gamma_{\lambda} = \gamma_{3}-\gamma_{2}=-uR_1/6, \quad
\gamma_{w} = \gamma_{2}-\gamma_{4}=uR_1/6-w, \nonumber \\
\gamma_{u} = 2\gamma_{6}-5\gamma_{3}/2-\gamma_{4}/2=
\left(4R_2-R_1\right)u-w/2,
\label{gammas2}
\end{eqnarray}
with the higher order corrections.

From the definitions (\ref{betagw}), relations (\ref{gammas2}) and explicit
expressions (\ref{gammaTwo}) for the anomalous dimensions we derive the
following leading-order expressions for the $\beta$ functions of our model:
\begin{equation}
\beta_{u} = u\, \left[-\varepsilon +Ru+ {w}/{2}\right],
\quad \beta_{w} = w\, \left[-\xi - {uR_1}/{6}+w\right],
\label{betas2}
\end{equation}
where we have introduced a new convenient parameter $R= R_1-4R_2$.

\section{Fixed points and scaling regimes} \label{sec:FPS}

It is well known that possible large-scale scaling regimes of a
renormalizable model are associated with IR attractive fixed points of the
corresponding RG equations. In our model, the coordinates $u_{*}$, $w_{*}$
of the fixed points are found from the equations
\begin{equation}
\beta_{u} (u_{*},w_{*}) = 0, \quad \beta_{w} (u_{*},w_{*})=0 ,
\label{points}
\end{equation}
with the $\beta$ functions given in (\ref{betagw}).
The type of a fixed point is determined by the matrix
\begin{equation}
\Omega=\{\Omega_{ij}=\partial\beta_{i}/\partial u_{j}\},
\label{OmegaDef}\nonumber
\end{equation}
where $\beta_{i}$ denotes the full set of the $\beta$ functions and
$u_{j}= \{u,w\}$ is the full set of couplings. For IR stable fixed points
the matrix $\Omega$ is positive, {\it i.e.}, the real parts of all its
eigenvalues are positive. This condition defines the regions of IR
stability for the corresponding scaling regimes.

The couplings $u$ and $w$ should be non-negative (by definition,
$u\propto g^2 \ge 0$ and $w \propto D_0/f\lambda \ge 0$), and
in the following we will be interested only in the
``good'' (admissible from the physics viewpoints) fixed points,
which  satisfy the conditions
\begin{equation}
u_{*}\ge 0, \quad w_{*}\ge 0
\label{posit}
\end{equation}
and can be IR attractive for some values of the model parameters.

In order to give the complete picture of possible scaling regimes,
it is instructive to discuss at first a more general case, specified
by the $\beta$ functions of the form
\begin{equation}
\beta_{u} = u [-\varepsilon+ au + bw], \quad
\beta_{w} = w [-\xi+ cu + dw],
\label{1}
\end{equation}
with arbitrary real coefficients $a$--$d$.

From equations (\ref{points}) and (\ref{1}) we can identify four different
fixed points. For the first three points the matrix $\Omega$ appears to be
triangular, so that its eigenvalues (and hence the regions of IR stability
of the corresponding scaling regimes) are simply determined by the diagonal
elements:
$\Omega_{u} = \partial \beta_{u} / \partial u >0$ and
$\Omega_{w} = \partial \beta_{w} / \partial w >0$.

\

1. Gaussian (free) fixed point: $u_{*}=w_{*}=0$;\ \
$\Omega_{u} = -\varepsilon$, $\Omega_{w} = -\xi$.

\

2. $u_{*}=0$ (exact result to all orders), $w_{*}=\xi/d$;
$\Omega_{u} = -\varepsilon+b\xi/d$, $\Omega_{w} = \xi$.
This point can be ``good'' only if $d>0$; otherwise the
the conditions $\Omega_{w}>0$ and $w_{*}>0$ cannot be simultaneously
satisfied.

\

3. $w_{*}=0$ (exact result to all orders), $u_{*}=\varepsilon/a$;
$\Omega_{u} = \varepsilon$, $\Omega_{w} = -\xi+c\varepsilon / a$.
Similarly to the case 2, this point can be ``good'' only if $a>0$.

\

The last, fully nontrivial, fixed point requires more detailed discussion.

\

4. The coordinates of this point are
\begin{equation}
u_{*}= (d\varepsilon - b\xi )/\Delta, \quad
w_{*}= (a\xi - c\varepsilon )/\Delta,
\quad \Delta = ad-bc,
\label{2}
\end{equation}
while the matrix $\Omega$ can be written in the form
\begin{eqnarray}
\Omega = \left(%
\begin{array}{cc}
  au_* &  bu_* \\&\\
  cw_* &  dw_* \\
\end{array}%
\right).
\label{omega4}
\end{eqnarray}
(it is useful not to substitute explicit expressions
(\ref{2}) into (\ref{omega4}) for a while).

The necessary and sufficient condition for the IR stability of
this point can be restated as the fulfillment of two inequalities:
\begin{equation}
\textrm{det}\, \Omega>0, \quad \textrm{tr} \,\Omega >0.
\label{5}\nonumber
\end{equation}
From (\ref{omega4}) one obtains
\begin{equation}
\textrm{det}\,\Omega = u_*w_*\Delta >0,
\label{6}\nonumber
\end{equation}
which along with (\ref{posit}) shows that this point can be ``good''
only if $\Delta >0$.

For the trace of $\Omega$ we obtain:
\begin{equation}
\textrm{tr} \Omega = au_* + dw_* >0.
\label{7}
\end{equation}
There are three possibilities:

1) $a>0$, $d>0$.
In this case the inequality (\ref{7}) is an automatic consequence of
(\ref{posit}). Four regions of
stability of the fixed points 1--4 divide the $\varepsilon$--$\xi$
plane without ``gaps'' or overlaps.
This is the most typical situation, realized for the $\varphi^4$ model or
the Gribov process in various kinds of random flows \cite{AHH}--\cite{AntM}.

2) $a<0, d<0$.
Then (\ref{7}) contradicts to (\ref{posit}) and this point can never be
``good.'' Such a situation was not yet encountered.

3) The parameters $a$ and $d$ are opposite in sign: $ad<0$.
For definiteness, we assume that $a<0$, $d>0$.
We will see in short that this situation can realize for the ATP model.
In this case one obtains from (\ref{7}):
\begin{equation}
w_* > -au_*/d >0,
\label{8}
\end{equation}
where the last inequality follows from $a<0$, $u_*>0$. The second inequality
in (\ref{posit}) is implied by the (\ref{8}) and thus becomes superfluous.
The region where the fixed point is IR attractive and positive is given by
the two inequalities
\begin{equation}
u_*>0, \quad au_*+dw_* >0.
\label{9}
\end{equation}

Summing up, we conclude that the scaling regime corresponding to the fixed
point 4 exists if $\Delta >0$ and at least one of the two parameters $a$
and $d$ is positive. The region where the fourth fixed point is ``good''
is determined by the inequalities (\ref{posit}) if $a$ and $d$ are
simultaneously positive, and by the conditions of the type (\ref{9})
if $a$ and $d$ are opposite in sign.

Let us turn to our specific model with the $\beta$ functions
(\ref{betas2}). Identifying them with (\ref{1}) gives
\begin{eqnarray}
a=R_1-4R_2\equiv R, \quad b=\frac{1}{2},\quad c=-\frac{R_1}{6}, \quad d=1;
\nonumber \\
\Delta=R+R_1/12 \equiv \Delta^R.
\label{ident}
\end{eqnarray}
The coordinates of the four possible fixed points are obtained by
substituting the expressions (\ref{ident}) into the general results.

\begin{figure}[p]
\abovecaptionskip=-7pt
\belowcaptionskip=16pt

\includegraphics[width=0.40\textwidth]{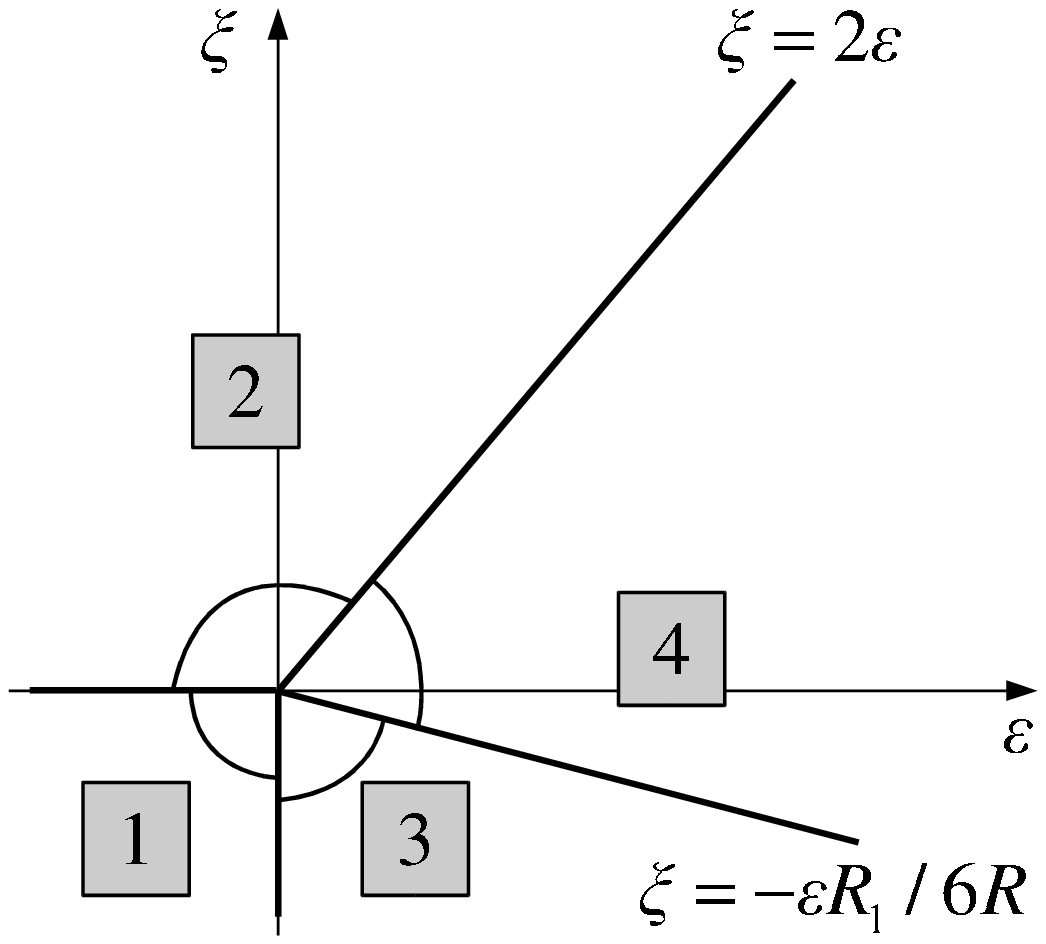}
{}\hfill{}
\includegraphics[width=0.40\textwidth]{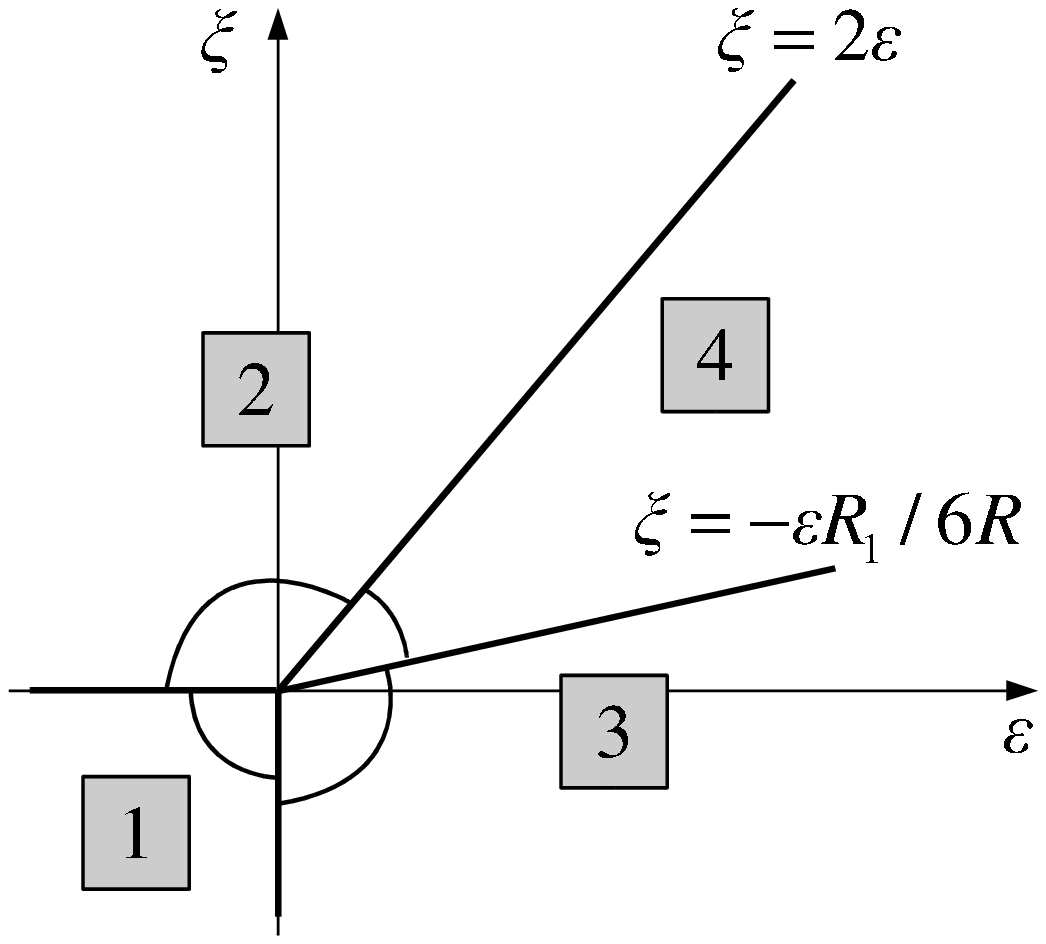}

\parbox[t]{0.45\textwidth}{\caption{ $\Delta^R>0$, $R>0$, $R_1>0$.}
\label{fig:pattern1}}
{}\hfill{}
\parbox[t]{0.45\textwidth}{\caption{ $\Delta^R>0$, $R>0$, $R_1<0$.}
\label{fig:pattern2}}

\includegraphics[width=0.40\textwidth]{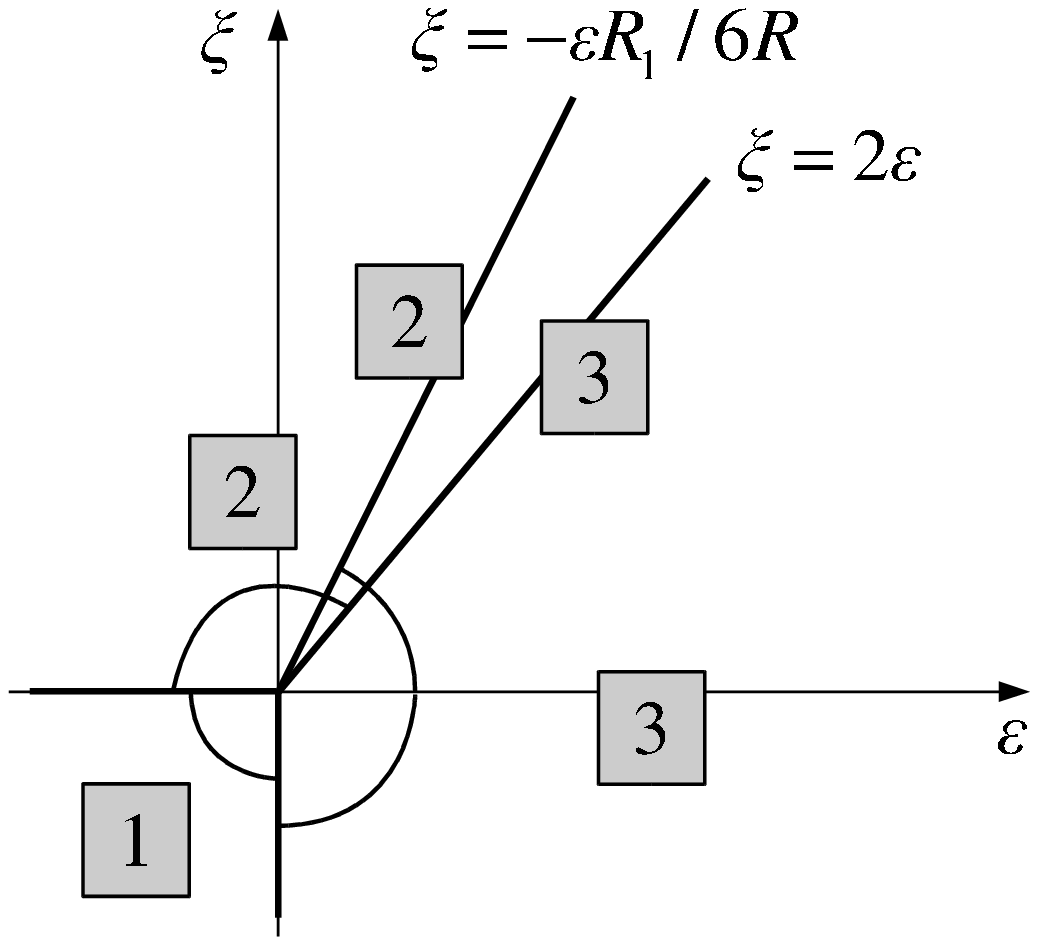}
{}\hfill{}
\includegraphics[width=0.40\textwidth]{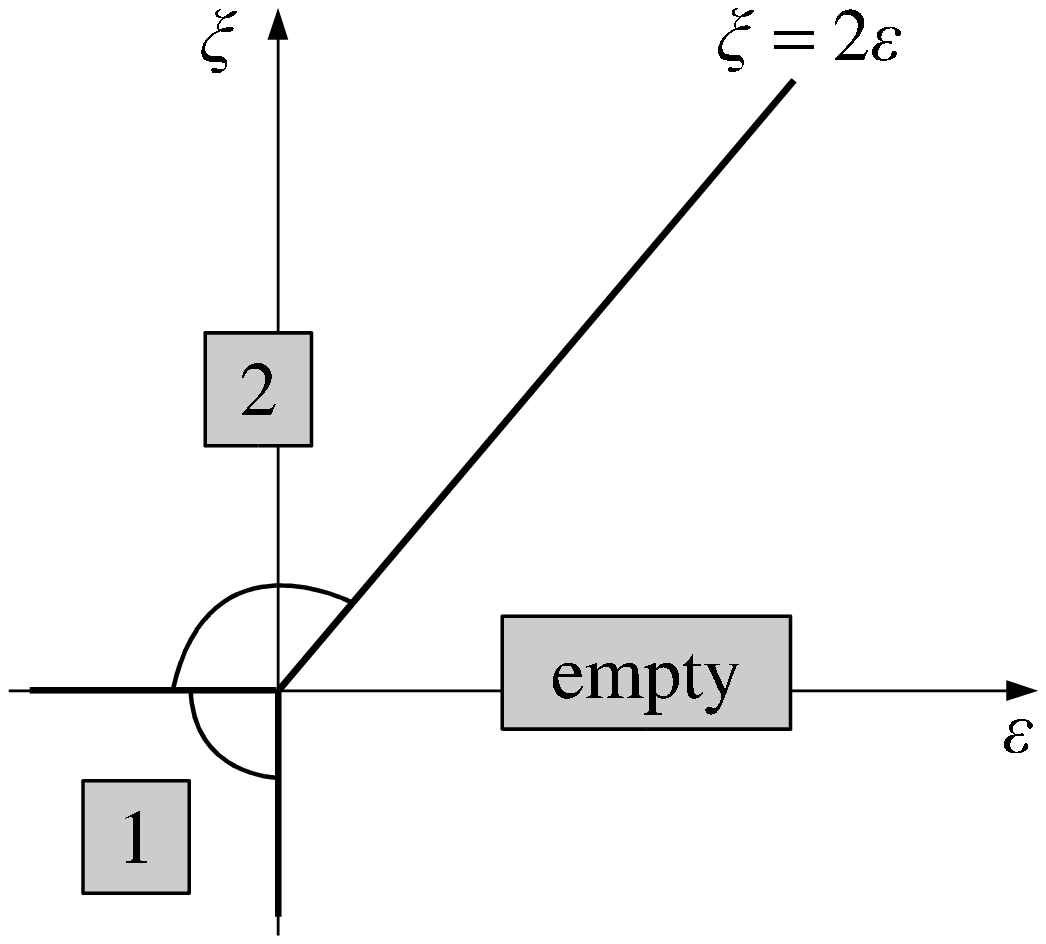}

\parbox[t]{0.45\textwidth}{\caption{ $\Delta^R<0$, $R>0$, $R_1<0$.}
\label{fig:pattern3}}
{}\hfill{}
\parbox[t]{0.45\textwidth}{\caption{ $\Delta^R<0$, $R<0$, $R_1$ arbitrary.}
\label{fig:pattern4}}
\belowcaptionskip=7pt
\begin{center}
\includegraphics[width=0.45\textwidth]{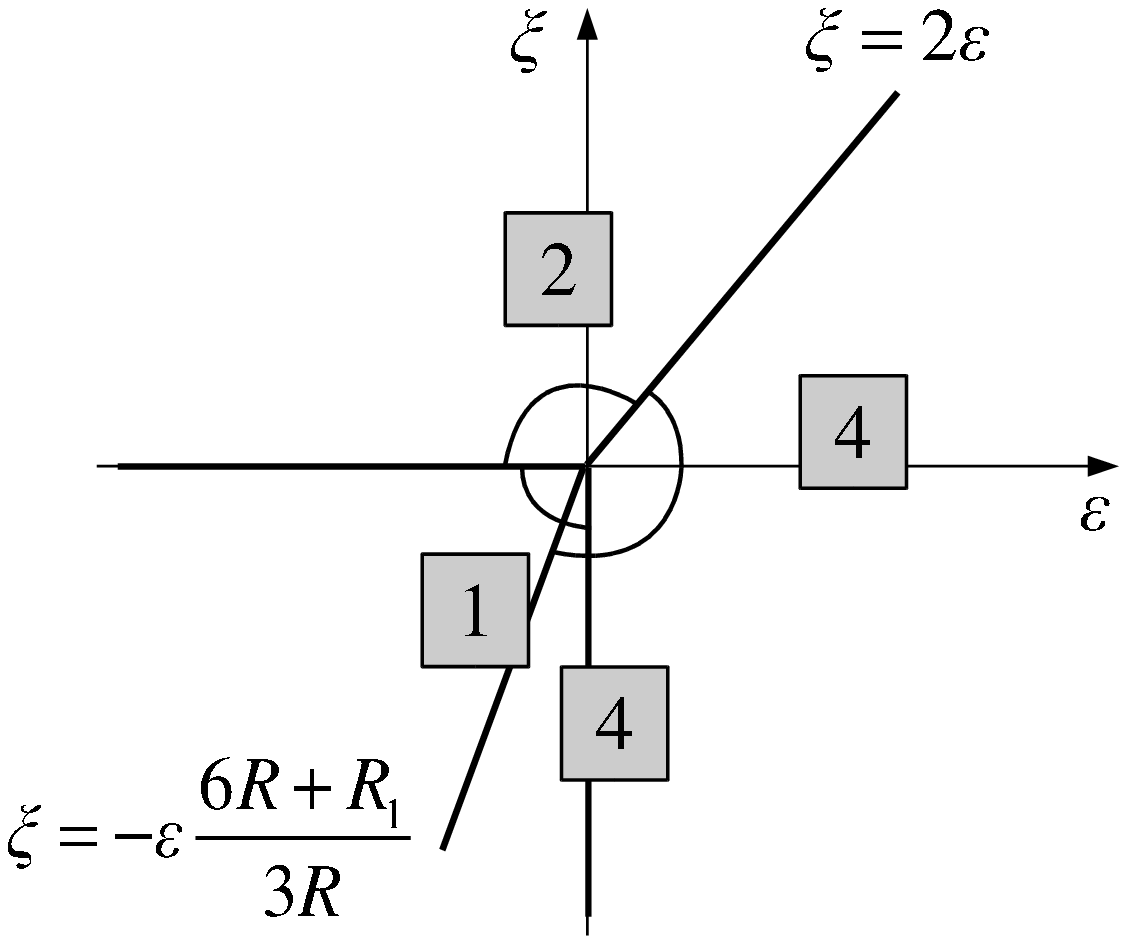}
\abovecaptionskip=5pt
\caption{\label{fig:pattern5}  $\Delta^R>0$, $R<0$, $R_1>0$.}
\end{center}

\begin{center}
\textbf{Regions of stability of
the fixed points in the model} (\protect\ref{action}).
\end{center}

\end{figure}

In the scaling regime corresponding to the fixed point~2, the nonlinearity
$\varphi^2$ in the stochastic equation (\ref{eq1}) becomes irrelevant in
the sense of Wilson due to the exact relation $u_{*}=0$. Thus we arrive at
the linear advection-diffusion equation for a passive scalar field $\varphi$.
In turn, the effects of the velocity field become irrelevant in the third
regime (fixed point~3) due to the exact relation $w_*=0$. The isotropy
violated by the velocity ensemble is restored and the leading terms of the
IR behaviour coincide with those of the equilibrium dynamic model ATP.
Finally, the fixed point 4 corresponds to a new nontrivial IR scaling regime,
in which the both nonlinearities in the stochastic equation for $\varphi$ are
important; the corresponding critical dimensions reveal strong anisotropy,
depend essentially on the both RG expansion parameters $\varepsilon$ and
$\xi$, and are calculated as double series in those parameters; see
section~\ref{sec:DimeNS}.

The regions of IR stability for all possible fixed points in the
$\varepsilon$--$\xi$ plane for different values of the parameters
$R_1$ and $R_2$ are shown on figures 1--5. In the one-loop approximation,
all the boundaries of the regions are given by straight lines.

The figures \ref{fig:pattern3} and \ref{fig:pattern5} show that there are
overlaps between the IR stability regions of fixed points~2 and~3 and
points~1 and~4, respectively. This means that for the values of $\varepsilon$
and $\xi$, corresponding to the region of an overlap, the system has two
variants of the IR scaling behaviour. Which one of them is realized
depends on the initial data for the parameters $u$ and $w$ in the RG
equations.

On the other hand, from figure \ref{fig:pattern4} one can see that if the
parameters $R_1$ and $R_2$ are such that  $\Delta_R<0$ and $R<0$, there is
a gap. The system does not exhibit scaling behaviour for corresponding values
of $\varepsilon$ and $\xi$, which can be interpreted as existence of a
first-order phase transition.

It is interesting to note that for $a<0$, the original static model has
no ``good'' fixed point (which is usually interpreted as a first-order
transition), but a ``good''  point of the type (\ref{2}) can appear
in the full dynamic model with two couplings, as illustrated by figure~5.
One can say that the phase transition changes its type and becomes a
second-order one owing to the turbulent mixing.

\begin{figure}

\includegraphics[width=0.45\textwidth]{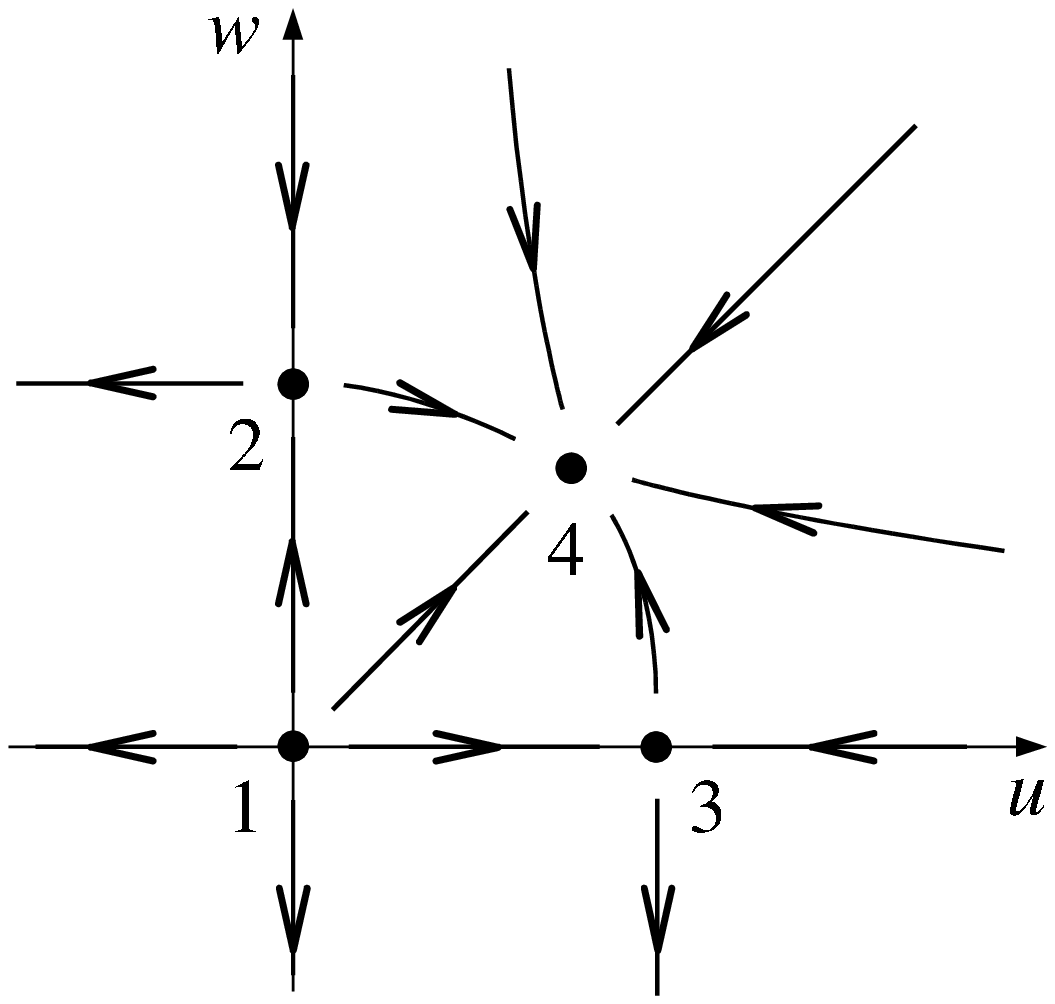}
{}\hfill{}
\includegraphics[width=0.45\textwidth]{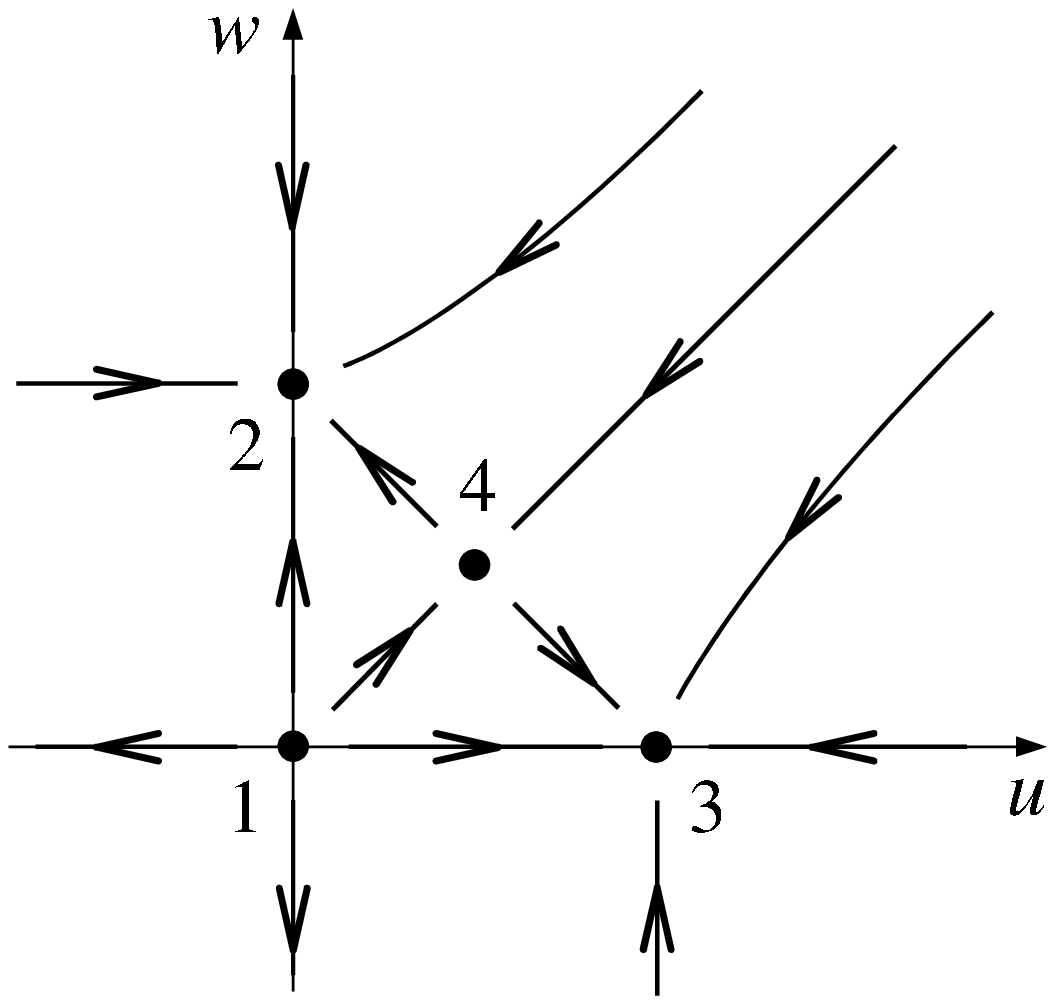}

\parbox[t]{0.45\textwidth}{\caption{RG flows for the case shown in figure~2.}
\label{fig:flows1}}
{}\hfill{}
\parbox[t]{0.45\textwidth}{\caption{RG flows for the case shown in figure~3.}
\label{fig:flows2}}

\end{figure}

Let us briefly discuss the pattern of the RG flows in the plane of couplings
$u$--$w$ for two special cases. Consider first the simplest (and the most
typical) situation, illustrated by figure~\ref{fig:pattern2}. The schematic
picture of the fixed points and RG flows is sketched on
figure~\ref{fig:flows1} for the case, when the values of the parameters
$\varepsilon$ and $\xi$ lie in the region~4 on figure~\ref{fig:pattern2}.
Then all the fixed points lie in the physical domain $u\ge0$, $w\ge0$: the
point~4 is IR attractive, point~1 is repulsive and points~2 and~3 are saddle
points. If the parameters $\varepsilon$ and $\xi$ are changing such that the
corresponding point in the $\varepsilon$--$\xi$ plane crosses the boundary
$\xi=2\varepsilon$ and moves into the region~2, the fixed point~4 on
figure~6 crosses the ray $u=0$, $w\ge0$, going through the point~2, and
moves into the unphysical domain $u<0$. The point~2 becomes IR attractive.
If the point in the figure~\ref{fig:pattern2} crosses the boundary
$\xi=-\varepsilon R_{1}/6R$, moving from region 4 to 3, the fixed point~4
crosses the ray $u\ge0$,
$w=0$, going through the point~3, and moves into the unphysical domain $w<0$;
the point~3 becomes IR attractive. Although this picture is based on the
explicit one-loop expressions for the $\beta$ functions and regions of
stability, it appears robust with respect to the higher-order corrections.
It is crucial here that the functions $\beta_{u}$ at $w=0$ and $\beta_{w}$ at
$u=0$ coincide with the $\beta$ functions of the corresponding single-charge
problems: the ATP model and the passive scalar case, respectively. The
latter are supposed to have a unique nontrivial fixed point each: for the
passive scalar, the $\beta$ function is known exactly, why for the ATP model
(with $R>0$) this is true at least within the $\varepsilon$ expansion.
Thus we may conclude that the coincidence of the fixed points~2 and~4 or~3
and~4 takes place simultaneously with the changeover in their type of
stability. In turn, this means that, although the boundaries between the
regions~2 and~4 or~3 and~4 in the $\varepsilon$--$\xi$ plane can become
curved beyond the one-loop approximation due to the higher-order
corrections to the $\beta$ functions in (\ref{1}), no gaps nor overlaps
can appear between those regions. This is equally true for the boundaries
between the regions~1 and~2 or~1 and~3 in figure~2, but there the boundaries
are not affected by the higher-order corrections due to the simple exact
expressions for the eigenvalues $\Omega_{u,w}$ at the Gaussian fixed point.

The RG flows for another interesting situation, illustrated by
figure~\ref{fig:pattern3}, are depicted on figure~\ref{fig:flows2}
for the case when the values of the parameters $\varepsilon$ and $\xi$
lie in the overlap of the regions~2 and~3. Then the both fixed
points~2 and~3 are ``good'' and the asymptotic behaviour of a flow
depends on the initial data for $u$ and $w$.
The point~4 is unphysical: although it lies in the domain $u\ge0$, $w\ge0$,
inspection of the explicit expressions (\ref{2}) and
(\ref{6}) shows that it is a saddle point for this case.
It is also worth noting that, for all possible situations, the RG flow with
the initial data in the physical domain $u\ge0$, $w\ge0$ can never leave it,
because the function $\beta_{u}$ vanishes for $u=0$ and arbitrary $w$, while
$\beta_{w}$ vanishes for $w=0$ and arbitrary $u$.

Let us conclude this section with a brief discussion of the most interesting
physical cases of the original ATP model. Then from (\ref{rr2}) and
(\ref{ident}) we obtain $R=(n+1)^{2} (7-3n)$ and
$\Delta^R= (n+1)^{2} (83-35n)/12$.

The case $n=2$ corresponds to the nematic-to-isotropic
transition in a liquid crystal. Then
\begin{eqnarray}
R=R_1=9,\quad R_2=0, \quad \Delta^R=39/4,
\label{N2}
\end{eqnarray}
and the case represented by figure \ref{fig:pattern1} is realized.
One can see that the interval of the most realistic values of the model
parameters, $d = 3$ and $0<\xi<2$, belongs completely to the region of
stability of the most nontrivial fixed point 4.

The second case of interest is the critical behaviour of bond percolation,
obtained in the limit $n=0$. Strictly speaking, more realistic
description of {\it dynamical} percolation is given by a special model with
a nonlocal in-time interaction \cite{JT}, but it is interesting to look at
the dynamics of the ATP model with $n=0$ as some kind of approximation.
For $n=0$ relations (\ref{rr2}) give
\begin{eqnarray}
R_1=-1,\quad R_2=-2,\quad R=7, \quad \Delta^R=83/12.
\label{N0}
\end{eqnarray}
Thus, we arrive at the case shown on figure \ref{fig:pattern2}. Again,
the most realistic values of the model parameters ($d=3$ and $\xi=4/3$)
belong to the stability region of the new anisotropic scaling regime,
corresponding to the fixed point 4.

For $n\ge3$ the case shown on figure~\ref{fig:pattern4} is realized. Now the
values $d=3$ and $\xi=4/3$ lie in the ``empty'' region where none of the
fixed points are ``good.''  This fact is usually interpreted as a first-order
transition; the turbulent mixing does not change its type. The interesting
situation illustrated by figure~\ref{fig:pattern5}, where the mixing gives
rise to the changeover in the type of the phase transition to the
second-order one, is realized for the interval $7/3 <n < 83/35$, which does
not contain integer values and has no documented physical interpretation.

\section{Critical scaling and critical dimensions} \label{sec:DimeNS}

We recall the definition of generalized homogeneity. Let $F$ be
a function of $n$ independent arguments
$\{ x_{1},... ,x_{n}\}$ that satisfies the dimensional relation
\begin{equation}
F(\lambda^{\alpha_{1}} x_{1}, \dots, \lambda^{\alpha_{n}} x_{n}) =
\lambda^{\alpha_{F}} F(x_{1}, \dots, x_{n})
\label{skal3}
\end{equation}
with a certain set of constant coefficients (scaling dimensions)
$\{ \alpha_{1}, \dots, \alpha_{n}, \alpha_{F} \}$ and an arbitrary positive
parameter $\lambda>0$. Differentiating relation (\ref{skal3}) with respect
to $\lambda $ and then setting $\lambda =1$, we obtain a
first-order differential equation with constant coefficients
\begin{eqnarray}
\sum_{i=1}^{n} \alpha_{i} {\cal D}_{i} \, F(x_{1}, \dots, x_{n}) =
\alpha_{F} \, F(x_{1}, \dots, x_{n}) , \quad {\cal D}_{i}=x_{i} \partial /
\partial x_{i}.
\label{skal1}
\end{eqnarray}
Its general solution has the form
\begin{eqnarray}
F(x_{1}, x_{2}, \dots, x_{n}) = x_{1}^{\alpha_{F}/\alpha_{1}} \,
\widetilde F \left( \frac{x_{2}}{x_{1}^{\alpha_{2}/\alpha_{1}}} \
,\dots,\ \frac{x_{n}} {x_{1}^{\alpha_{n}/\alpha_{1}}} \right),
\label{skal2}
\nonumber
\end{eqnarray}
where $\widetilde F$ is an arbitrary function of $(n-1)$ arguments.
Obviously, the dimensions are defined up to a common
factor (this can be seen by replacing $\lambda\to\lambda^{a}$
in (\ref{skal3}) or multiplying equation (\ref{skal1}) by $a$); this
arbitrariness can
be eliminated, for example, if we set $\alpha_{1}=1$. If $\alpha_{i}=0$
for some $x_{i}$, this variable is not dilated
in (\ref{skal3}), and the corresponding derivative is absent
from (\ref{skal1}).

It is well known that the leading terms, determining the asymptotic
behaviour of (renormalized) correlation functions at large distances,
satisfy the RG equation (\ref{RG1}), in which the renormalized coupling
constants are replaced with their values at the fixed points. In our case,
this leads to
the equation
\begin{eqnarray}
\left\{ D_{\mu} - \gamma_{f}^{*} D_{f} - \gamma_{\lambda}^{*} D_{\lambda}
-  \gamma_{\tau}^{*} D_{\tau} +\gamma^{*} _{\varphi} N_{\varphi}+\gamma^{*}
_{\varphi'} N_{\varphi'}  \right\}
\,G^{R} =0,
\label{RGE}
\end{eqnarray}
where $\gamma^{*} = \gamma (u=u_{*}, w=w_{*})$ for all the anomalous
dimensions.

We are interested in the critical scaling behaviour, that is, behaviour of
the type (\ref{skal3}) in which all the IR relevant parameters
(momenta/coordinates, frequencies/times, deviation of the temperature
from its critical value $\tau \propto (T-T_{c})$) are dilated, while the
IR irrelevant parameters (those which remain finite at the fixed point:
$\lambda$, $\mu$ and $f$) are fixed \cite{Zinn,Book3}. Thus we combine the
equation (\ref{RGE}) with the analogous equations, corresponding to the
canonical scale invariance (see section \ref{sec:Reno}), so that the
derivatives with respect to the IR irrelevant parameters are eliminated;
this gives the desired equation which describes the critical scaling
behaviour (for more details, see {\it e.g.} \cite{Alexa,AntM}):
\begin{eqnarray}
\left\{ D_{\bot} +\Delta_{\parallel} D_{\parallel} +
\Delta_{\omega} D_{\omega} + \Delta_{\tau} D_{\tau} - N_{\Phi}
\Delta_{\Phi} \right\} G_{N_{\Phi}} =0,
\label{Krit}\nonumber
\end{eqnarray}
where $D_{\bot} = k_{\bot} \partial / \partial k_{\bot}$ and
$D_{\parallel} = k_{\parallel} \partial / \partial k_{\parallel}$.
Here, $\Delta_{\bot}=1$ is the normalization condition, and the critical
dimension of any IR-relevant parameter $F$ is
given by the general expression
\begin{eqnarray}
\Delta_{F} = d_{F}^{\bot}+ \Delta_{\parallel} d_{F}^{\parallel}+
\Delta_{\omega} d_{F}^{\omega} + \gamma^{*}_{F},
\label{KritDim}
\end{eqnarray}
with canonical dimensions from Table \ref{table1} and the relations
\begin{eqnarray}
\Delta_{\omega}=2-\gamma_{\lambda}^{*} , \quad
\Delta_{\parallel} = \left( 2+\gamma_{f}^{*} \right)/2 .
\label{KritDim2}
\end{eqnarray}

We are in a position to write the final one-loop results for the critical
dimensions. Substituting (\ref{gammas2}) into the general formulae
(\ref{KritDim}) and (\ref{KritDim2}) gives
\begin{eqnarray}
\Delta_{\omega} = 2+\frac{u_*R_1}{6}, \quad
\Delta_{\parallel}=1+\frac{w_*}{2},\quad
\Delta_{\tau} = 2+\frac{5u_*R_1}{3}
\nonumber\\
\Delta_{\varphi'} = \frac{d}{2}+1+\frac{u_*R_1}{3}+\frac{w_*}{4}, \quad
\Delta_\varphi=\frac{d}{2}-1+\frac{u_*R_1}{6}+\frac{w^*}{4}.
\nonumber
\end{eqnarray}
By inserting the explicit expressions for the fixed point coordinates and
taking the equality $d = 6 -\varepsilon$ into account one obtains the
leading-order expressions for the critical dimensions. The results for all
scaling regimes are summarized in Table~2.

\begin{table}
\caption{Critical dimensions of the fields and parameters in the model
(\protect\ref{action})}
\label{table3}
\begin{tabular}{|c|c|c|c|c|}
\hline
 & FP1 & FP2 & FP3 &  FP4  \\
\hline
$\Delta_{\omega}$ & 2 & 2 & $2+\frac {R_1\varepsilon } {6R}$ &
$2+\frac{R_1\left(2\varepsilon-\xi\right)}{12\Delta^R}$ \\
\hline
$\Delta_{\parallel}$ & 1 & $1+\frac \xi 2$ & 1 &
$1+\frac{6R\xi+R_1\varepsilon}{12\Delta^R}$ \\
\hline
$\Delta_{\tau}$ & 2 & 2 & $2+\frac{5R_1\varepsilon}{3R}$ &
$2+\frac{5R_1\left(2\varepsilon-\xi\right)}{6\Delta^R}$ \\
\hline
$\Delta_{\varphi'}$ & $4-\frac \varepsilon 2$ &
$4-\frac \varepsilon 2+\frac \xi 4$ &
$4+\frac {\left(2R_1-3R\right)\varepsilon} {6R}$ &
$4+\frac{\left(9R_1-12\Delta^R\right)\varepsilon+
\left(6R-4R_1\right)\xi} {24\Delta^R}$ \\
\hline
$\Delta_{\varphi}$ & $2-\frac \varepsilon 2$ & $2-\frac \varepsilon 2
+\frac \xi 4$ & $2+\frac {\left(R_1-3R\right)\varepsilon} {6R}  $ &
$2+\frac{\left(5R_1-12\Delta^R\right)\varepsilon+\left(6R+2R_1\right)\xi}
{24\Delta^R}$  \\
\hline
\end{tabular}
\end{table}
The expressions for the first and second fixed points are exact.
Other dimensions have corrections, given by higher powers of $\varepsilon$
for the third fixed point and higher powers of $\varepsilon$ and $\xi$ for
the fourth one.
The critical dimensions for the models of a liquid crystal and bond
percolation are derived from the general results by substituting
the expressions (\ref{N2}) and (\ref{N0}), respectively.

Let us discuss the consequences of the general scaling relations for the most
interesting special case of the pair correlation function. They result in the
scaling expression
\begin{eqnarray}
\big\langle \varphi_{a} ({\bf x}+{\bf r}, t+t')\, \varphi_{b} ({\bf x}, t')
\big\rangle = \delta_{ab}
r_{\bot}^{-2\Delta_{\varphi}} \,
{\cal F} \left( \tau_{0}\,{r^{\Delta_{\tau}}}, \,
t/r_{\bot}^{\Delta_{\omega}}, \,
r_{\parallel}/ r_{\bot}^{\Delta_{\parallel}} \right),
\label{scaling1}
\end{eqnarray}
where  $r_{\bot}=|{\bf r}_{\bot}|$, $r_{\parallel}=|{\bf r}_{\parallel}|$ \
and ${\cal F}$ is some scaling function. This representation is valid in
the symmetric phase ($\tau_{0} \ge0$), where the tensor
structure is simply given by the $\delta$ symbol. It is natural to assume
that ${\cal F}$ has a finite limit for $\tau_{0}=0$ (that is, exactly at
the critical point) and/or for $t=0$ (equal-time correlation function).
Then from (\ref{scaling1}) one obtains
\begin{eqnarray}
\big\langle \varphi_{a}({\bf x}+{\bf r},t)\, \varphi_{b}({\bf x},t)
\big\rangle = \delta_{ab}
 r_{\bot}^{-2\Delta_{\varphi}} \, \widetilde{\cal F}
\left( r_{\parallel}/ r_{\bot}^{\Delta_{\parallel}} \right)
\label{scaling2}\nonumber
\end{eqnarray}
with another nontrivial function $\widetilde{\cal F}(x)={\cal F}(0,0,x)$.

The two last arguments in the scaling representation (\ref{scaling1}) can
also be chosen in the form $r_{\bot}/ L_{\bot} (t)$ and $r_{\parallel}
/L_{\parallel} (t)$ with two different characteristic length scales
\begin{eqnarray}
L_{\bot} (t) \sim t^{\alpha_{\bot}}, \quad
L_{\parallel} (t) \sim t^{\alpha_{\parallel}}, \qquad
\alpha_{\bot} = 1/\Delta_{\omega}, \quad
\alpha_{\parallel} = \Delta_{\parallel}/\Delta_{\omega}.
\label{scales}
\end{eqnarray}
For the most realistic values $\varepsilon=3$ ($d=3$) and $\xi=4/3$
(Kolmogorov spectrum of the velocity) and for the $n=2$ case of our model
(liquid crystals) the explicit results from Table~2 and the expressions
(\ref{N2}) give
\begin{eqnarray}
\Delta_{\omega} \approx 2.359, \quad \Delta_{\parallel}
\approx 1.846, \quad \alpha_{\bot} \approx 0.424,  \quad
\alpha_{\parallel} \approx 0.783, \quad  \Delta_{\varphi} \approx 1.487,
\label{scales2}
\nonumber
\end{eqnarray}
while for the percolation limit $n=0$ from (\ref{N0})
one obtains
\begin{eqnarray}
\Delta_{\omega} \approx 1.944, \quad \Delta_{\parallel}
\approx 1.639, \quad \alpha_{\bot} \approx 0.514,
\quad \alpha_{\parallel} \approx 0.843, \quad
\Delta_{\varphi} \approx 0.731.
\label{scales3}
\nonumber
\end{eqnarray}
Existence of two different length scales (\ref{scales}) with power-law
dependence on the time was established earlier in a number of studies
within numerical simulations \cite{Shear2}, approximate analytical solutions
\cite{Shear3}, RG analysis \cite{Alexa,AntM} and exactly soluble simplified
models \cite{Shear1}. It is interesting to note that the inequality
$\alpha_{\parallel} > \alpha_{\bot}$ also holds for all those cases.

\section{Conclusion} \label{sec:Conc}

We studied effects of turbulent mixing on the critical behaviour of
the system, described by a relaxational dynamics of a non-conserved
order parameter of the ATP model. The mixing was modelled by a Gaussian
statistics with vanishing correlation time and strongly anisotropic
correlation function $\propto\delta(t-t') /k_{\bot}^{d-1+\xi}$; see
equations (\ref{veloc1}), (\ref{veloc2}). Such ensembles were employed
earlier in \cite{AM}--\cite{AM2} in the analysis of the two-dimensional
passive turbulent advection (linear equation for the scalar field).

The model, originally described by stochastic differential equations
(\ref{eq1})--(\ref{LG}), (\ref{nabla}), can be reformulated as a
multiplicatively renormalizable field theory with the action (\ref{action}),
which allows one to employ the field theoretic RG to study its
critical behaviour. The model reveals four
different IR scaling regimes, related with the four different fixed
points of the RG equations. Their regions of stability
in the $\varepsilon$--$\xi$
plane were identified in the leading order of the double expansion in
$\varepsilon$ and $\xi$ and are shown on figures 1--5.
These regimes correspond to:

(1) Gaussian (free) model;

(2) Linear passive scalar advection
(the self-action term in the ATP Hamiltonian (\ref{LG}) is irrelevant
in the sense of Wilson);

(3) Equilibrium critical dynamics of the ATP model (interaction with
the velocity field is irrelevant); and

(4) The full-fledged strongly anisotropic scaling regime in which the both
interactions are important; it corresponds to a new non-equilibrium
universality class.

It was shown that the equilibrium critical regimes for the both physically
interesting cases (liquid crystals and percolation process) become
unstable for the realistic range of parameters $d=3$ and $0<\xi\le 2$,
which includes the Kolmogorov spectrum ($\xi=4/3$) and the Batchelor limit
($\xi\to 2$) and is replaced with the new non-equilibrium regime.
The corresponding critical dimensions
were calculated to first order of the corresponding RG expansion, which
in this case takes on the form of the double expansion in
$\varepsilon$ and $\xi$; explicit expressions are given in Table~2.

Those results were derived within the leading (one-loop) approximation,
that is, in the leading order of the double expansion in $\varepsilon$ and
$\xi$, and their validity for finite physical values of these parameters
can be called in question (especially because of large physical values of
$\varepsilon=6-d$). Careful analysis of this problem requires calculation
of the higher-order corrections and applying some kind of summation
procedure to the results obtained, as was done {\it e.g.} in \cite{Bonfim}
for the scalar static $\varphi^{3}$ model. Such analysis goes far beyond
the scope of the present paper, and we hope to address it in the future.
Nevertheless, the discussion of the RG flows, given in section~\ref{sec:FPS},
suggests that the pattern of fixed points (and thus of critical regimes),
obtained within the one-loop approximation, appears robust with respect
to higher-order corrections and is preserved for finite values of
$\varepsilon$ and $\xi$.

Thus we hope that our simplified model of a non-conserved order parameter
and Gaussian velocity ensemble captures the most important features of the
full-fledged problem: emergence of a new non-equilibrium universality
class with a new set of critical exponents, completely different from those
of the classical ATP model; existence (for a strongly anisotropic velocity
ensemble) of two different length scales (with a power law time dependence),
and so on. Further investigation should take into account conservation of
the order parameter, compressibility of the fluid, non-Gaussian character
and finite correlation time of the velocity field, and so on.
This work is partly in progress and partly remains for the future.

\section*{Acknowledgments}

The authors thank L~Ts Adzhemyan, Michal Hnatich, Juha Honkonen and
Paolo Muratore Ginanneschi for discussions. AVM was supported in part
by the Dynasty Foundation.

\end{document}